\documentstyle[12pt,pra,epsf,aps]{revtex}
\begin{document}
\title{On the Floquet Theory of Delay Differential Equations}
\author{C.~Simmendinger, A.~Wunderlin}
\address{Institut f\"ur Theoretische Physik und Synergetik,
Universit\"at Stuttgart,\\
Pfaffenwaldring 57/4, D-70550 Stuttgart, Germany}
\author{A.~Pelster}
\address{Institut f\"ur Theoretische Physik, Freie Universit\"at Berlin,\\
Arnimallee 14, D-14195 Berlin, Germany}
\date{\today}
\maketitle
\begin{abstract}
We present an analytical approach to deal with
nonlinear delay differential equations close to instabilities of time periodic reference states.
To this end we start with approximately
determining such reference states by extending the
Poincar\'e Lindstedt and the Shohat expansions which were originally
developed for ordinary differential equations.
Then we systematically elaborate a linear stability analysis around a
time periodic reference state. This
allows to approximately calculate the Floquet eigenvalues and
their corresponding eigensolutions
by using matrix valued continued fractions. \\

Pacs: 02.30.Ks, 05.45.+b
\end{abstract}

\section{Introduction}

Over the last two decades
considerable new interest in the theory
of delay differential equations
led to various remarkable results \cite{hale,diek,wo,mike}.
The reason is that the solution space for delay differential equations
has to be considered as infinite dimensional although only a finite number
of dynamical variables is involved \cite{krasovskii}. As a consequence,
nonlinear delay differential equations reveal a broad class of
instabilities leading from oscillatory to chaotic behavior. Apart from
the period doubling route to chaos also quasi periodic states, intermittency
and locking behavior have been observed in detailed numerical studies
\cite{mike}. In particular in the chaotic domain it has been suggested
that the envelope to the Kaplan Yorke dimension of a delay induced
chaotic attractor is proportional to the time delay
\cite{mike,farmer,berre,pom}.
This fact offers the possibility to generate high dimensional chaotic
attractors by simply increasing the time delay. \\

Delay differential equations have been successfully applied to model
numerous nonlinear systems where dynamical instabilities are induced by
the finite propagation time of signals in feedback loops. For instance,
experiments on optical devices, acusto optic and electro optic
bistable devices \cite{ikeda,lako,petermann} have confirmed
both the theoretical
and the numerical predictions. But
delay induced instabilities play also an important role in other disciplines
such as population dynamics \cite{hale},
radio engineering sciences \cite{kislov},
economy \cite{bellmann} and biology\cite{heiden}.
In addition it has been noted in medical sciences
that there exists a
remarkable variety of clinically relevant dynamical phenomena
under physiological and pathological conditions.
For example oscillations or chaotic behavior can spontaneously occur or
disappear as a function of external or internal time delays as has been
demonstrated by the Mackey Glass model of blood circulation \cite{mackey},
the Cheyne Stokes
respiration \cite{glass}, and the forearm tracking with visual
delayed feedback \cite{wunderlin}. \\

The interesting properties of nonlinear
delay differential equations
have been mainly investigated in numerical studies.
Therefore it becomes desirable to
substantiate these results - at least in comparably simple situations -
by analytical methods. An interesting result
in this direction has been recently obtained in
\cite{wo} by
rigorously analyzing the instability of
a time independent reference state. The application
of the theory to a delay induced
Hopf bifurcation has been
confirmed by numerical as well as experimental studies.
We note, that a different method, which is based upon a multiple
scaling analysis, has been recently demonstrated in \cite{giapo}.

Here, however, it is our aim to
generalize the approach of \cite{wo} by starting from a time periodic reference
state and by analytically investigating its stability.\\

Our paper is organized as follows. In Section II we introduce two methods
for approximately determining a time periodic reference state. Section III
then develops its linear stability analysis. The resulting Floquet
theory leads to a homogeneous vector valued recurrence relation determining
the Floquet eigenvalues and its corresponding eigensolutions. In Section IV
we offer two solution methods for this recurrence relation which are based
on matrix valued continued fractions. Section V completes the Floquet theory
by studying the adjoint problem. Eventually Section VI is devoted to a
short summary and several conclusions in view of possible future work.

For a numerical derivation of the Floquet exponents and the 
corresponding eigenvectors for this case we refer to \cite{berre1,berre2}. 

\section{Determination of the time periodic reference state}

We assume that the dynamical behavior of the system under consideration can
be characterized by
a state vector $\vec q(t)$ in an $n$ dimensional state space
 $\Gamma$ and that the underlying equation of motion is an autonomous
delay differential equation of the general form
\begin{eqnarray}
\label{nonlineveq}
\frac{d}{dt}\vec q(t)=\vec N(\vec q(t),\vec q(t-\tau),\{\sigma_i\}).
\end{eqnarray}
Here $\vec N$ denotes a nonlinear vector field which depends on the state
vector $\vec q$ at the
times $t$ and $t-\tau$, respectively, with $\tau$ representing  the time delay.
The set $\{\sigma_i\}$ describes the control parameters which measure
external influences on the system. We assume that these
control parameters are kept fixed so that we can omit them in our notation.\\

The treatment of the nonlinear
problem (\ref{nonlineveq}) close to an instability strongly depends on the
chosen reference
state. A theory for an instability of a time independent reference state
has been recently developed in \cite{wo}.
 Here it is our aim to generalize this method to situations
 where we start from a time periodic reference state.
 As such states cannot be expressed by closed analytical forms,
it becomes necessary to describe them by using proper approximation
schemes. In this section we extend methods which have been developed
in the realm of ordinary differential equations towards delay differential
equations. The Poincar\'e Lindstedt
approximation allows to determine the time periodic reference state for
small values of a parameter, whereas its improvement, the Shohat method
turns out to possess a wider range of applicabilities
\cite{shohat,minorsky}. \\

\subsection{The Poincar\'e Lindstedt expansion}

A perturbative approximation method, as the Poincar\'e Lindstedt expansion,
relies on the existence of a suitable smallness parameter $\mu$. Dealing
with the nonlinear differential equation (\ref{nonlineveq}), we have to
distinguish in general two different origins for such a smallness parameter
$\mu$. On the one hand, the smallness parameter $\mu$ can be generated by a
delay induced instability. Then it measures the relative deviation of the
time delay $\tau$ from the critical value $\tau_c$ above which the delay
induced time periodic reference state exists. This case occurs, for instance,
in the electronic phase locked loop with time delay \cite{wo} where
the underlying model equation reveals a Hopf bifurcation at some $\tau_c$.
Considering the corresponding normal form \cite{manneville}
\begin{eqnarray}
\frac{dZ}{dt}=\sigma Z -g|Z|^2 Z
\end{eqnarray}
we may choose
$\mu = \sqrt{(\tau - \tau_c)/\tau_c}$. On the other hand, the smallness
parameter $\mu$ can also coincide with one of the given control parameters
of the system. An example is provided by a harmonic oscillator with the
frequency $\omega_0$ which is driven by a nonlinear time delayed perturbation:
\begin{eqnarray}
\label{plnl}
q''(t,\mu) +\omega_0^2 q(t,\mu) = \mu
f(q(t,\mu),q'(t,\mu),q(t-\tau,\mu),q'(t-\tau,\mu)).
\end{eqnarray}
Here $q(t,\mu)$ denotes a scalar variable,
the prime abbreviates the derivative with respect to the time $t$ and
$f$ represents a nonlinear function of its arguments.\\

For the sake of
simplicity we now discuss the Poincar\'e Lindstedt expansion
not for the general delay differential equation (\ref{nonlineveq}) but
only for the model equation (\ref{plnl}).
We start with the situation
of a vanishing smallness parameter $\mu$ where the solution of (\ref{plnl})
is a periodic reference state
\begin{eqnarray}
q(t,0)=q(t+T_0,0) ,
\end{eqnarray}
with $T_0=2\pi/\omega_0$ denoting the period of the unperturbed oscillator.
Switching on the smallness parameter $\mu$, this state will be transformed
to a new periodic state, which can be described by
\begin{eqnarray}
q(t,\mu)=q\left(t+\frac{2\pi}{\omega(\mu)},\mu\right) .
\end{eqnarray}
In the following it becomes useful to explicitly take into account the
frequency shift from $\omega_0$ to $\omega(\mu)$ by rescaling
the time $t$ according to
\begin{eqnarray}
\label{umsk1}
\xi(t)& = & \omega(\mu) t .
\end{eqnarray}
Introducing the new variable
\begin{eqnarray}
\label{umsk2}
x(\xi,\mu) & = & q\left(\frac{\xi}{\omega(\mu)},\mu\right),
\end{eqnarray}
which is $2\pi$ periodic in $\xi$,
\begin{eqnarray}
\label{perio}
x(\xi,\mu)=x(\xi+2\pi,\mu) ,
\end{eqnarray}
we can rewrite the equation of motion (\ref{plnl}) as:
\begin{eqnarray}
\label{umbe2}
\omega(\mu)^2 \ddot x(\xi,\mu) +\omega_0^2 x(\xi,\mu) = \mu f(x(\xi,\mu),
\dot x(\xi,\mu),x(\xi-\omega(\mu)\tau,\mu), \dot x(\xi-\omega(\mu)\tau,\mu)).
\end{eqnarray}
The dot indicates the derivative with respect to the dimensionless
new time variable $\xi$.\\

As already mentioned, we assume that $\mu$ represents a small quantity
so that we can
expand the frequency
$\omega(\mu)$ and the periodic orbit $x(\xi,\mu)$ in powers of $\mu$
according to
\begin{eqnarray}
\label{entwi1}
x(\xi,\mu) & = & x_0(\xi)+\mu x_1(\xi)+\mu^2 x_2(\xi)+\cdots ,\\
\label{entwi2}
\omega(\mu) & = & \omega_0+\mu\omega_1+\mu^2\omega_2+\cdots .
\end{eqnarray}
In addition to the similar procedure for ordinary differential equations
\cite{shohat,minorsky}
we have also to consider a corresponding
expansion of the time delayed terms in (\ref{umbe2}).
This is achieved by
\begin{eqnarray}
\label{entwi3}
x(\xi -\omega(\mu)\tau,\mu) & = &
x_0(\xi-\omega_0\tau) + \mu\left(x_1(\xi-\omega_0\tau)-\omega_1\tau\dot
x_0(\xi-\omega_0\tau)\right)+\cdots , \\
\label{entwi4}
\dot x(\xi -\omega(\mu)\tau,\mu) & = &  \dot
x_0(\xi-\omega_0\tau) + \mu\left(\dot
x_1(\xi-\omega_0\tau)-\omega_1\tau\ddot
x_0(\xi-\omega_0\tau)\right)+\cdots .
\end{eqnarray}
If we apply these expansions to the equation of motion (\ref{umbe2}) and
combine terms of the same power of $\mu$, we obtain in each order a system
of inhomogeneous linear ordinary differential equations of second order:
\begin{eqnarray}
\ddot x_0(\xi)+x_0(\xi) & = & 0 ,\\
\ddot x_1(\xi)+x_1(\xi) & = & -2\frac{\omega_1}{\omega_0}\ddot x_0(\xi)
+\frac{1}{\omega_0^2}f(x_0(\xi),\dot x_0(\xi),
x_0(\xi-\omega_0\tau),\dot x_0(\xi-\omega_0\tau)) ,\\
\vdots & = & \vdots \nonumber\\
\label{in}
\ddot x_n(\xi)+x_n(\xi) & = & I_n(\xi) .
\end{eqnarray}
The inhomogeneity $I_n(\xi)$ which appears in the $n$th order (\ref{in})
is purely determined by the lower order terms $x_m(\xi),  0\leq m< n$.
We have to guarantee that our periodicity condition (\ref{perio}) is
fulfilled in each order of the perturbation theory. However, if the
Fourier expansion of the inhomogeneity $I_n(\xi)$ includes multiples of
the first harmonic terms which are proportional to $\sin(\xi)$ or
$\cos(\xi)$, the solution $x_n(\xi)$ of (\ref{in}) contains aperiodic
secular terms of the form $\xi\sin(\xi)$ or $\xi\cos(\xi)$, respectively.
We can avoid these aperiodic solutions by demanding
\begin{eqnarray}
\label{ford1}
\int^{2\pi}_0 I_n(\xi)\sin(\xi) d\xi = 0 , \quad
\int^{2\pi}_0 I_n(\xi)\cos(\xi) d\xi = 0 .
\end{eqnarray}
In order to fulfill these two conditions we need two independent
parameters. Here we choose the constant $\omega_n$ as the first parameter,
whereas the second one can be chosen by imposing suitable initial
conditions for
$x_{n-1}(\xi)$, for example
\begin{eqnarray}
\label{anf1}
x_{n-1}(0) = A_{n-1} , \quad
\dot x_{n-1}(0) = 0 .
\end{eqnarray}
In this way we obtain a systematic approximation scheme to determine our time
periodic reference state order by order for small values of the
parameter $\mu$.

\subsection{The Shohat expansion}

In a situation where the parameter $\mu$ is not a small quantity the
Poincar\'e Lindstedt expansion for the calculation of the
time periodic reference
state has to be modified. This can be achieved by introducing a new smallness
parameter $\rho(\mu)$ by the prescription
\begin{eqnarray}
\label{shohaent}
\rho(\mu)=\frac{\mu}{1+\mu}
\end{eqnarray}
which maps the interval $[0,\infty)$ of $\mu$ onto the interval $[0,1)$
of $\rho$.
The resulting method of the Shohat expansion can be described as follows.
The equation of motion (\ref{umbe2}) is multiplied by $\mu^2$.
In doing so we become able to expand the periodic reference state $x(\xi,\mu)$
as well as the product $\mu\omega(\mu)$ with respect to $\rho$  and obtain
\begin{eqnarray}
\label{omeg0}
x(\xi,\mu) & = & X_0(\xi)+\rho(\mu) X_1(\xi)+\rho(\mu)^2
X_2(\xi)+\cdots ,\\
\label{omeg1}
\mu\omega(\mu) & = &
\rho(\mu)\Omega_0+\rho(\mu)^2\Omega_1+\rho(\mu)^3\Omega_2+\cdots .
\end{eqnarray}
In order to guarantee that the frequency $\omega(\mu)$ approaches the
frequency $\omega_0$ of the unperturbed harmonic oscillator in the limit
$\mu\to 0$
we have to choose $\Omega_0=\omega_0$.
>From (\ref{omeg0}) and (\ref{omeg1}) and the inversion of the
relation (\ref{shohaent})
\begin{eqnarray}
\label{omeg2}
\mu &= & \frac{\rho(\mu)}{1-\rho(\mu)}
\end{eqnarray}
we deduce the expansions
\begin{eqnarray}
\omega(\mu) & = & \Omega_0+\rho(\mu)\left(\Omega_1
-\Omega_0\right)+\rho(\mu)^2\left(\Omega_2-\Omega_1\right)+\cdots ,\\
\label{omeg3}
x(\xi-\omega(\mu)\tau,\mu) & = & X_0(\xi-\Omega_0\tau)+\nonumber\\ & &
+\rho(\mu)\left[X_1(\xi-\Omega_0\tau)-(\Omega_1-\Omega_0)\tau\dot
X_0(\xi-\Omega_0\tau)\right]+\cdots .
\end{eqnarray}
The further application of the Shohat method is completely analogous
to the Poincar\'e Lindstedt approximation scheme.
It has been conjectured without proof \cite{shohat}, that the method works for
arbitrary parameter values $\mu\ge 0$. We thus note, that although to our knowledge no known counterexample for this conjecture exists, the vailidity of this
expansion has to be confirmed for each case individually.

\section{Stability of the time periodic reference state}

We now generalize the linear stability analysis of a time independent
reference state developed in \cite{wo} to the case of a time periodic
reference state.
To that end we return to the general form of the delay differential
equation (\ref{nonlineveq}) and rescale the time according to
(\ref{umsk1}),(\ref{umsk2}):
\begin{eqnarray}
\label{umskaliert}
\frac{d}{d\xi}\vec q(\xi) & = & \frac{1}{\omega}\vec
N(\vec q(\xi),\vec q(\xi-\omega\tau),\{\sigma_i\}).
\end{eqnarray}
Here $\omega=\omega(\mu)$ abbreviates the frequency of the
time periodic reference state, henceforth denoted by $\vec q^{\,0}(\xi)
=\vec q^{\,0}(\xi,\mu)$.\\

Following the original notion of  Krasovskii and Hale \cite{hale,krasovskii}
as well as its detailed elaboration in \cite{wo} we generalize the $n$
dimensional state space $\Gamma$ to an infinite dimensional state space
${\cal C}$.
This allows to embed the given delay differential equation (\ref{umskaliert})
in the context of functional differential equations.
It turns out that this reformulation represents the adequate framework
for a linear stability analysis around a time periodic reference state.
The resulting Floquet theory leads to a homogeneous
vector valued recurrence relation
which determines the Floquet eigenvalues as well as the corresponding Floquet
eigensolutions.

\subsection{Formulation of the problem in the extended state space}

It appears that solutions of the delay differential equation
(\ref{umskaliert}) for times $\xi\geq 0$ depend on initial values of the
state vector $\vec q(\xi)$ in the entire interval $[-\omega\tau,0]$.
Therefore we have to complete (\ref{umskaliert}) with the initial condition
\begin{eqnarray}
\label{initcon}
\vec q(\theta)=\vec g(\theta),\quad -\omega\tau\leq\theta\leq 0,
\end{eqnarray}
where $\vec g$ is a given continuous vector valued function in a suitable
function space ${\cal C}$.
The initial value problem (\ref{umskaliert}), (\ref{initcon})
then maps the function $\vec g$  onto a trajectory in the $n$ dimensional
state space $\Gamma$. Therefore the problem arises that different
initial vector valued functions  $\vec g$ may yield crossings of the
corresponding trajectories in $\Gamma$. This means that the point-wise
uniqueness of solutions can not be assured when we restrict our
considerations to the state space $\Gamma$.\\

In order to solve this problem one may introduce the extension of the finite
dimensional state space  $\Gamma$ to an infinite dimensional function
space $\cal C$ where the initial vector valued function $\vec g$ is defined.
According to Krasovskii and Hale \cite{hale,krasovskii} this is achieved
by regarding the trajectory $\vec q(\xi)$ in  the original state space
$\Gamma$ during the time interval $[\xi-\omega\tau,\xi]$ as a single point
$\vec q_\xi$ in the extended space $\cal C$ (compare Fig. 1):
\begin{eqnarray}
\label{tdef}
\vec q_\xi(\theta)=\vec q(\xi+\theta),\quad -\omega\tau\leq\theta\leq 0.
\end{eqnarray}
The dynamics of the delay system can then also be described in the extended
state space $\cal C$  by introducing the nonlinear solution
operator ${\cal T}(\xi)$:
\begin{eqnarray}
\label{solop}
\vec q_{\xi}(\theta)=\left({\cal T}(\xi)\vec g\right)(\theta),
\quad -\omega\tau\leq\theta\leq 0 .
\end{eqnarray}
Its uniqueness is expressed by the fact that the operator ${\cal T}(\xi)$
has the properties of a semi group, that is
\begin{eqnarray}
{\cal T}(\xi+\eta)={\cal T}(\xi){\cal T}(\eta),\quad \xi,\eta\geq 0,
\quad {\cal T}(0)={\cal I},
\end{eqnarray}
where ${\cal I}$ denotes the identity operator. We now have to reformulate the
original initial value problem (\ref{umskaliert}), (\ref{initcon})
in the extended space $\cal C$. To this end we formally differentiate
(\ref{solop}) with respect to the time $\xi$
\begin{eqnarray}
\label{infingen0}
\frac{d}{d\xi}\vec q_{\xi}(\theta)=\left({\cal A}\vec
q_{\xi}\right)(\theta),\quad -\omega\tau\leq\theta\leq 0 .
\end{eqnarray}
Here $\cal A$ denotes  the infinitesimal generator which corresponds to the
solution operator ${\cal T}(\xi)$:
\begin{eqnarray}
\label{gren}
\left({\cal A}\vec q_{\xi}\right)(\theta)=\lim_{\epsilon \to 0}
\frac{1}{\epsilon}\left[\left({\cal T}(\epsilon)\vec q_\xi\right)
(\theta)-\vec q_\xi(\theta)\right] .
\end{eqnarray}
By evaluating this limit separately for the interval
$-\omega\tau\leq\theta < 0$ and
for the point $\theta=0$ we obtain the explicit expression \cite{wo}
\begin{eqnarray}
\label{nlin0}
\left({\cal A}\vec q_\xi\right)(\theta)=\left\{\begin{array}{ll}
\displaystyle\frac{d}{d\theta}\vec q_\xi(\theta) , &
\quad -\omega\tau\leq\theta <0 , \\ \displaystyle
{\cal N}[\vec q_\xi(.)],  & \quad \theta=0. \end{array} \right.
\end{eqnarray}
The nonlinear functional $\cal N$ is constructed as follows. We assume that the
original vector field $\vec N$ in (\ref{umskaliert}) can be expanded into
powers of its arguments $\vec q(\xi)$ and $\vec q(\xi-\omega\tau)$.
A typical term of second order in this expansion has, for instance, the form
\begin{eqnarray}
N_{ijk}^{(2)}q_j(\xi)q_k(\xi-\omega\tau) ,
\end{eqnarray}
where the explicit components of the
respective vectors have been introduced and summation
is understood over dummy indices. The representation of $\vec q(\xi)$ and
$\vec q(\xi-\omega\tau)$
can be given in terms of the extended state space $\cal C$ by taking into
account the relation (\ref{tdef}):
\begin{eqnarray}
\vec q(\xi) = \int^0_{-\omega\tau} d\theta\delta(\theta)
\vec q_\xi(\theta) , \quad
\vec q(\xi-\omega\tau) =  \int
^0_{-\omega\tau}d\theta\delta(\theta+\omega\tau)\vec q_\xi(\theta) .
\end{eqnarray}
If we apply this procedure to every term in the series expansion and collect
terms of the same order in the extended state vector $\vec q_\xi$, the
nonlinear vector
field $\vec N$ becomes a vector valued functional ${\cal N}$ with the
components
\begin{eqnarray}
{\cal N}_i[\vec q_\xi(.)] & = & \sum^{\infty}_{k=1}
\int^0_{-\omega\tau}d\theta_1\dots\int^0_{-\omega\tau}d\theta_k
\frac{1}{\omega}{\Omega}^{(k)}_{i,j_1\dots j_k}(\theta_1,
\dots,\theta_k)q_{\xi,j_1}(\theta_1)\dots
q_{\xi,j_k}(\theta_k),
\end{eqnarray}
where the ${\Omega}^{(k)}_{i,j_1\dots j_k}(\theta_1,\dots,\theta_k)$
represent matrix valued densities.
Thus we have reached our first goal, namely to derive a nonlinear functional
differential equation for the problem formulated in (\ref{umskaliert}),
(\ref{initcon}).

\subsection{The linearized equation of motion}

According to the prescription (\ref{tdef}) the time periodic reference state
$\vec q^{\,0}(\xi)$ in the state space $\Gamma$ transforms into
$\vec q^{\,0}_\xi(\theta)$ in the extended state space ${\cal C}$.
In order to test its linear stability we insert the ansatz
\begin{eqnarray}
\label{linear}
\vec q_\xi(\theta)=\vec q^{\,0}_\xi(\theta)+\tilde{\vec q_\xi}(\theta)
\end{eqnarray}
into (\ref{infingen0}), (\ref{nlin0}).
Dropping the tilde we obtain in the linear approximation for the
infinitesimal deviation $\vec q_{\xi}(\theta)$
\begin{eqnarray}
\label{levol}
\frac{d}{d\xi}\vec q_{\xi}(\theta) & = & \left({\cal A}_L \vec
q_{\xi}\right)(\theta) ,
\end{eqnarray}
where the linear infinitesimal generator ${\cal A}_L$ becomes explicitly
time dependent:
\begin{eqnarray}
\label{ansb}
\left({\cal A}_L \vec q_{\xi}\right)(\theta) & = &
\left\{\begin{array}{ll}\displaystyle \frac{d}{d\theta}\vec
q_{\xi}(\theta), & \quad -\omega\tau\leq\theta<0 ,\\
\displaystyle\int^0_{-\omega\tau}d\theta'
{\bf \Omega}_{\xi}(\theta')\vec q_{\xi}(\theta'), & \quad\theta=0 .
\end{array}\right.
\end{eqnarray}
The matrix valued density ${\bf \Omega}_{\xi}(\theta)$ can be written
as a functional derivative of  $\cal N$ evaluated at the time periodic
reference state $\vec q_\xi^{\,0}$:
\begin{eqnarray}
\label{omdef}
{\bf \Omega}_{\xi}(\theta) = \left[\frac{\delta{\cal
N}[\vec q_\xi(.)]}{\delta \vec q_\xi(\theta)}\right]_{\vec q_\xi=\vec
q_\xi^{\,0}} .
\end{eqnarray}

\subsection{Transformation of the linear problem}

Due to the fact that the reference state $\vec q_\xi^{\,0}(\theta)$ is
$2\pi$ periodic with respect to $\xi$, the matrix valued density
${\bf \Omega}_{\xi}(\theta)$ in  (\ref{omdef}) is time dependent
with the same period. We thus perform a Fourier
expansion of the matrix ${\bf \Omega}_{\xi}(\theta)$ :
\begin{eqnarray}
\label{fourmeg}
{\bf \Omega}_{\xi}(\theta) & = & \sum^\infty_{k=-\infty}{\bf
\Omega}_k(\theta)e^{i k\xi} .
\end{eqnarray}
In close analogy to the Floquet theorem for ordinary differential
equations \cite{haken2} we try to solve (\ref{levol})--(\ref{fourmeg})
by the ansatz
\begin{eqnarray}
\label{ansatz0}
\vec q_\xi(\theta) & = & e^{\lambda\xi}\vec\phi^\lambda_{\xi}(\theta).
\end{eqnarray}
Here  $\lambda$ denotes the Floquet eigenvalue and
$\vec\phi^\lambda_\xi(\theta)=\vec\phi^\lambda_{\xi+2\pi}(\theta)$
is a $2\pi$ periodic Floquet eigensolution for which we also perform
a Fourier expansion:
\begin{eqnarray}
\label{ansatz}
\vec \phi^\lambda_\xi(\theta)=\sum^{\infty}_{n=-\infty}
\vec\phi^{\lambda}_n(\theta)e^{i n\xi} .
\end{eqnarray}
In order to determine the Fourier components  $\vec\phi^\lambda_n(\theta)$
we insert the hypothesis (\ref{ansatz0}), (\ref{ansatz}) into the linearized
equation of motion (\ref{levol})--(\ref{fourmeg}).
We now have to consider separately the interval  $-\omega\tau\leq\theta<0$
and the point $\theta=0$.
In the interval $-\omega\tau\leq\theta<0$ we conclude that the
Fourier component $\vec\phi^{\lambda}_n(\theta)$ has the form
\begin{eqnarray}
\vec\phi^{\lambda}_n(\theta) & = & \vec\phi^{\lambda}_n
e^{(\lambda+ in)\theta} .
\end{eqnarray}
For the case  $\theta=0$ we find
\begin{eqnarray}
\label{fouriercomp}
\sum^\infty_{n=-\infty}\vec\phi^\lambda_n(\lambda+in)
e^{(\lambda+i n)\xi}=
\sum^\infty_{n=-\infty}\sum^\infty_{k=-\infty}
\int^0_{-\omega\tau}d\theta \, {\bf
\Omega}_k(\theta)e^{(\lambda+in)\theta}e^{(i (k+n)+\lambda)\xi}
\vec\phi^\lambda_n .
\end{eqnarray}
We now introduce the matrix valued quantity
\begin{eqnarray}
\label{ldefin}
{{\bf L}}_{k,n} & = & \int^0_{-\omega\tau}d\theta
\, {\bf \Omega}_{k}(\theta)e^{(\lambda+i n)\theta}
\end{eqnarray}
and a new index  $\tilde n(n)=n+k$.
Comparing the contributions of the various Fourier components
in (\ref{fouriercomp}) and dropping the tilde, we obtain a homogeneous
vector valued recurrence relation for the Fourier components
$\vec\phi^\lambda_{n}$:
\begin{eqnarray}
\label{rekurs}
0 & = & \sum^\infty_{k=-\infty}\left[{{\bf L}}_{k,
n-k}-\delta_{k,0}(\lambda+i n){\bf I}\right]\vec\phi^\lambda_{n-k} .
\end{eqnarray}
Thus we are left with the problem to construct an approximate solution to
(\ref{rekurs}) which leads to both the Floquet eigenvalues $\lambda$ and
the corresponding Floquet eigensolutions.

\subsection{Remark}

In the Floquet theory of ordinary differential equations \cite{haken2}
it is shown that the derivative of the time periodic reference state
represents a Floquet eigensolution where the real part of the
corresponding Floquet eigenvalue vanishes. This statement remains valid
for delay differential equations as can be seen as follows. As the
time periodic reference state $\vec q^{\,0}_\xi(\theta)$ satisfies the
nonlinear equation of motion (\ref{infingen0}), (\ref{nlin0}),
a differentiation with respect to the time $\xi$ leads to
\begin{eqnarray}
\label{PS}
\frac{d}{d \xi} \frac{d \vec q_{\xi}^{\,0} (\theta)}{d \xi} & = &
\left\{\begin{array}{ll}\displaystyle \frac{d}{d\theta}
\frac{d \vec q_{\xi}^{\,0} (\theta)}{d \xi} ,
& \quad -\omega\tau\leq\theta<0 ,\\
\displaystyle\int^0_{-\omega\tau}d\theta'  \, \left[
\frac{\delta{\cal N}[\vec q_\xi(.)]}{\delta
\vec q_\xi(\theta')}\right]_{\vec q_\xi=\vec q_\xi^{\,0}}
\frac{d \vec q_{\xi}^{\,0} (\theta')}{d \xi} , & \quad\theta=0 .
\end{array}\right.
\end{eqnarray}
A comparison with (\ref{levol})--(\ref{omdef}) reveals that the derivative
of the time periodic reference state $\vec q^{\,0}_\xi(\theta)$
indeed fulfills the linear problem. Due to (\ref{ansatz0}) and (\ref{ansatz})
it therefore possesses the general form
\begin{eqnarray}
\label{ressu}
\frac{d \vec q^{\,0}_\xi (\theta)}{d \xi} = e^{\lambda \xi}
\sum^{\infty}_{n=-\infty} \vec \phi^{\lambda}_n(\theta) e^{i n\xi} .
\end{eqnarray}
>From the $2 \pi$ periodicity of $\vec q^{\,0}_\xi(\theta)$ and its
derivative (\ref{ressu}) we the conclude that the real part of its
Floquet eigenvalue $\lambda$ has to vanish.

\section{Matrix valued continued fractions}

We consider two methods which enable us to approximately solve
(\ref{rekurs}) for the Floquet eigenvalues
$\lambda$ and for the Fourier components $\vec\phi^\lambda_n$ of the
Floquet eigensolutions. In the first part we formulate a new method
based on $n$ diagonal continued fractions.
In the second part we show that this solution method is equivalent to a
formulation with tridiagonal continued fractions introduced by Risken
\cite{risken}. It turns out, however, that the first method is much
simpler to handle
as the necessary inversion of matrices can be performed in a
low dimensional space. Furthermore the criterion for truncating
higher order  terms in the smallness parameter $\mu$ can
be formulated more precisely in the framework of the first method.

\subsection{Pentadiagonal recurrence relations}

We apply the method of matrix valued continued fractions in order to solve
the vector valued recurrence relation (\ref{rekurs}) approximately.
In order to avoid overloading the notation we restrict ourselves
for the time being to
the pentadiagonal case where the summation in (\ref{rekurs}) is
performed for $-2\leq k\leq 2$:
\begin{eqnarray}
\label{ndiag}
0  =   {{\bf L}}_{-2,n+2}\vec\phi^\lambda_{n+2}+{{\bf
L}}_{-1,n+1}\vec\phi^\lambda_{n+1}+[{{\bf
L}}_{0,n}-(\lambda+i n){\bf
I}]\vec\phi^\lambda_n + {{\bf
L}}_{1,n-1}\vec\phi^\lambda_{n-1}+{{\bf
L}}_{2,n-2}\vec\phi^\lambda_{n-2} .
\end{eqnarray}
We start with defining a set of ladder operators ${\bf S}_n^m$
for $m=\pm 1,\pm 2$ which relate neighboring Fourier components via
\begin{eqnarray}
\label{s1}
\vec\phi^\lambda_{n+m} & = & {\bf S}^{m}_n\vec\phi^\lambda_n .
\end{eqnarray}
This definition implies the following useful relations between different
ladder operators:
\begin{eqnarray}
{\bf S}_n^{-1} = \left[{\bf S}_{n-1}^{+1}\right]^{-1},\quad\quad
{\bf S}^{+1}_{n+1}{\bf S}^{+1}_n & = & {\bf S}_n^{+2} .
\end{eqnarray}
Applying the definition (\ref{s1}) of the ladder operators, the
pentadiagonal recurrence relation (\ref{ndiag}) can be rewritten as:
\begin{eqnarray}
\label{tran}
0  = \left[{{\bf L}}_{-2,n+2}{\bf S}^{+2}_n+{{\bf L}}_{-1,n+1}{\bf
S}^{+1}_n+[{{\bf L}}_{0,n}-(\lambda+i n){\bf
I}] + {{\bf L}}_{1,n-1}{\bf
S}^{-1}_n+{{\bf L}}_{2,n-2}{\bf S}^{-2}_n\right]\vec\phi^\lambda_{n} .
\end{eqnarray}
We now express the ladder operators  ${\bf S}^m_n (m=\pm 1 ,\pm 2)$
in terms of  the matrices  ${\bf L}_{k,n}$ as well as the operators
${\bf S}^m_{n\pm1},{\bf S}^m_{n\pm2}$.
In order to evaluate this dependence, we isolate the term
$\vec\phi^\lambda_{n+1}={\bf S}^{+1}_n\vec\phi^\lambda_n$ in (\ref{tran}).
Then the
equation assumes the form
\begin{eqnarray}
\label{isol}
{\bf L}_{-1,n+1}\vec\phi^\lambda_{n+1}  =  -\left[ \left[{\bf
L}_{0,n}-(\lambda+i n){\bf I}\right]+{\bf L}_{-2,n+2}{\bf
S}^{+2}_n + {\bf L}_{1,n-1}{\bf S}^{-1}_n+{\bf
L}_{2,n-2}{\bf S}^{-2}_n\right]\vec\phi^\lambda_n  .
\end{eqnarray}
Shifting the index from $n$ to $n-1$ and applying the definition
$\vec\phi^\lambda_{n-1}  =  {\bf S}^{-1}_n\vec\phi^\lambda_n$
we obtain from the validity for all $\vec\phi^\lambda_{n}$ the operator
relation
\begin{eqnarray}
\label{s1m}
{\bf S}_n^{-1}  =  -\left[ \left[{\bf
L}_{0,{n-1}}-(\lambda+i (n-1)){\bf I}\right]+{\bf
L}_{-2,n+1}{\bf S}^{+2}_{n-1} + {\bf L}_{1,n-2}{\bf
S}^{-1}_{n-1}+{\bf L}_{2,n-3}{\bf S}^{-2}_{n-1}\right]^{-1}{\bf
L}_{-1,n} .
\end{eqnarray}
Similarly we construct the operator relations
\begin{eqnarray}
\label{s1p}
{\bf S}_n^{+1} = -\left[ \left[{\bf
L}_{0,{n+1}}-(\lambda+i (n+1)){\bf I}\right]+{\bf
L}_{-2,n+3}{\bf S}^{+2}_{n+1}+{\bf
L}_{-1,n+2}{\bf S}^{+1}_{n+1}+{\bf L}_{2,n-1}{\bf
S}^{-2}_{n+1}\right]^{-1}{\bf L}_{1,n} , \\
{\bf S}_n^{-2}  =  -\left[ \left[{\bf
L}_{0,{n-2}}-(\lambda+i (n-2)){\bf I}\right]+{\bf
L}_{-1,n-1}{\bf S}^{+1}_{n-2} + {\bf L}_{1,n-3}{\bf
S}^{-1}_{n-2}+{\bf L}_{2,n-4}{\bf S}^{-2}_{n-2}\right]^{-1}{\bf
L}_{-2,n} ,\\
\label{s2m}
{\bf S}_n^{+2}  =  -\left[ \left[{\bf
L}_{0,{n+2}}-(\lambda+i (n+2)){\bf I}\right]+{\bf
L}_{-1,n+3}{\bf S}^{+1}_{n+2} + {\bf
L}_{1,n+1}{\bf S}^{-1}_{n+2}+{\bf L}_{-2,n+4}{\bf
S}^{+2}_{n+2}\right]^{-1}{\bf L}_{2,n} .
\end{eqnarray}
We perform an iteration procedure by starting from (\ref{tran})
for the case $n=0$,
\begin{eqnarray}
\label{trake}
0 = \left[{{\bf L}}_{-2,2}{\bf S}^{+2}_0+{{\bf L}}_{-1,1}{\bf
S}^{+1}_0+[{{\bf L}}_{0,0}-\lambda{\bf I}]+
{{\bf L}}_{1,-1}{\bf S}^{-1}_0+{{\bf L}}_{2,-2}{\bf
S}^{-2}_0\right]\vec\phi^\lambda_{0},
\end{eqnarray}
and by recursively inserting the recurrence relations of the ladder
operators (\ref{s1m})-(\ref{s2m}).
Writing the successive inversions of the matrices formally as fractions,
we may visualize this iteration procedure by a schematic representation
of a pentadiagonal matrix valued continued fraction.
This is illustrated in Fig. 2 where each term is represented by
a horizontal line and where bifurcations are related to the terms of the
ladder operators ${\bf S}^m_n (m=\pm 1 ,\pm 2)$.\\

This far we discussed the solution of the vector valued recurrence
relation (\ref{tran}) in the pentadiagonal case for $m=\pm 1, \pm 2$.
However,  our method can be correspondingly  extended  to  the general
case where all Fourier components $\vec\phi^\lambda_{n}$ are coupled to
each other.
To this end we introduce ladder operators ${\bf S}^m_n$ with arbitrary $m$
according to (\ref{s1}), where we identify ${\bf S}^0_n={\bf I}$.
The homogeneous vector valued recurrence relation (\ref{rekurs}) then
yields a corresponding one for the ladder operators ${\bf S}^m_n$:
\begin{eqnarray}
\label{rekurssmn}
0 & = & \sum^\infty_{k=-\infty}\left[{{\bf L}}_{k,
n-k}-\delta_{k,0}(\lambda+i n){\bf I}\right]{\bf S}^{-k}_n .
\end{eqnarray}
An iteration procedure similar to (\ref{s1m})--(\ref{s2m}) finally leads to a
homogeneous equation for the Fourier component $\vec\phi^\lambda_0$,
\begin{eqnarray}
\label{homogeneous}
{\bf M}(\lambda)\vec\phi^\lambda_0 = 0,
\end{eqnarray}
where the resulting matrix ${\bf M}(\lambda)$ consists of an infinite
number of matrix valued continued fractions transcendentally depending
on the Floquet eigenvalues $\lambda$. Therefore the Floquet eigenvalues
$\lambda$ are determined
from the condition that the determinant of the matrix ${\bf M}(\lambda)$
vanishes:
\begin{eqnarray}
\label{cond}
\det {\bf M}(\lambda) = 0.
\end{eqnarray}
Once the Floquet eigenvalues $\lambda$ are known, the yet unknown
Fourier component $\vec\phi^\lambda_0$ is determined up to a constant
from solving the homogeneous equation (\ref{homogeneous}).
All Fourier components $\vec\phi_n^\lambda$ of the Floquet eigensolutions
are then calculated by successively applying the ladder operators
${\bf S}^m_n$ starting with $\vec\phi^\lambda_0$:
\begin{eqnarray}
\vec\phi^\lambda_n= {\bf S}^n_0\vec\phi^\lambda_0.
\end{eqnarray}
Note that the remaining normalization constant in $\vec\phi^\lambda_0$
has to be fixed
by an adequate biorthonormality condition which will be discussed in
Section V.D.\\

In applications, however, it is impossible to exactly evaluate the
infinite number of matrix valued continued fractions.
>From an analytical point of view we can therefore expect that this solution
method allows at most to approximately determine Floquet eigenvalues and the
corresponding eigensolutions.
To this end we recall that the starting point of our linear stability analysis,
 i.e. the time periodic reference state, is only known as a finite power series
in the smallness parameter $\mu$.
As a consequence the whole calculation can be simplified by approximately
neglecting higher order terms in the smallness parameter $\mu$.
In particular it becomes sufficient to restrict the vector valued recurrence
relation (\ref{rekurs}) to a finite number of terms, so that the subsequent
iteration procedure only leads to finite number of matrix valued continued
fractions. Furthermore each continued fraction can be evaluated in the
leading order of the smallness parameter $\mu$. In spite of these
successive expansions,
the continued fractions have the property that the approximate results
rapidly converge towards the exact values if the leading order in the
smallness parameter $\mu$ is increased.

\subsection{Sketch of Risken's tridiagonal formulation}

Following Risken \cite{risken} we show that the $n$ diagonal matrix valued
continued fractions can always be cast into a tridiagonal form.
For the sake of simplicity we demonstrate this only for the pentadiagonal
recurrence relation (\ref{ndiag}), but the general case is treated along
similar lines. We start with distinguishing between even and odd indices
$n$ in the pentadiagonal recurrence relation (\ref{ndiag}):
\begin{eqnarray}
0 & =& {{\bf L}}_{-2,2n+2}\vec\phi^\lambda_{2n+2}+{{\bf L}}_{-1,2n+1}
\vec\phi^\lambda_{2n+1}+ \nonumber \\
\label{ungege}
& & +[{{\bf L}}_{0,2n}-(\lambda+i 2n){\bf I}]
\vec\phi^\lambda_{2n}+{{\bf L}}_{1,2n-1}\vec\phi^\lambda_{2n-1}
+{{\bf L}}_{2,2n-2}\vec\phi^\lambda_{2n-2}, \\
0 & = & {{\bf L}}_{-2,2n+3}\vec\phi^\lambda_{2n+3}+{{\bf L}}_{-1,2n+2}
\vec\phi^\lambda_{2n+2}+ \nonumber\\
\label{ungegr}
& & +[{{\bf L}}_{0,2n+1}-(\lambda+i (2n+1))
{\bf I}]\vec\phi^\lambda_{2n+1}+{{\bf L}}_{1,2n}\vec\phi^\lambda_{2n}
+{{\bf L}}_{2,2n-1}\vec\phi^\lambda_{2n-1} .
\end{eqnarray}
We now construct new vectors according to the prescription
\begin{eqnarray}
\vec \Phi^\lambda_{n+1}  =  \left({\vec\phi^\lambda_{2n+2}\atop
\vec\phi^\lambda_{2n+3}}\right), \quad
\vec\Phi^\lambda_{n}  =  \left({\vec\phi^\lambda_{2n}\atop\vec
\phi^\lambda_{2n+1}}\right), \quad
\vec\Phi^\lambda_{n-1}  =  \left({\vec\phi^\lambda_{2n-2}\atop
\vec\phi^\lambda_{2n-1}}\right) .
\end{eqnarray}
Additionally we define the matrices
\begin{eqnarray}
{\bf Q}_{-1,n+1} = \left(\begin{array}{cc} {{\bf L}}_{-2,2n+2}  0\\
{{\bf L}}_{-1,2n+2}  {{\bf
L}}_{-2,2n+3}\end{array}\right)\nonumber , \quad
{\bf Q}_{1,n-1}  =  \left(\begin{array}{cc} {{\bf L}}_{2,2n-2}
{{\bf L}}_{1,2n-1}\\ 0  {{\bf L}}_{2,2n-1}\end{array}\right) , \\
{\bf Q}_{0,n}  = \left(\begin{array}{cc} [{{\bf
L}}_{0,2n}-(\lambda+2i n){\bf I}]  {{\bf L}}_{-1,2n+1}\\
{{\bf L}}_{1,2n}  [{{\bf
L}}_{0,2n+1}-(\lambda+i(2n+1)){\bf I}]
\end{array}\right). \hspace*{1cm}
\end{eqnarray}
Due to these definitions both pentadiagonal recurrence relations
(\ref{ungege}), (\ref{ungegr}) can be combined in the following way:
\begin{eqnarray}
\label{triris}
{\bf Q}_{-1,n+1}\vec\Phi^\lambda_{n+1}+{\bf
Q}_{0,n}\vec\Phi^\lambda_{n}+{\bf Q}_{+1,n-1}\vec\Phi^\lambda_{n-1}=0 .
\end{eqnarray}
Thus we have reached our goal to find a tridiagonal vector valued recurrence
relation.
At this stage we define again ladder operators ${\bf R}^\pm_n$ with the
property
\begin{eqnarray}
\label{sbest1}
\vec \Phi^\lambda_{n+1}  =  {\bf R}^{+}_n\vec\Phi^\lambda_n , \quad
\vec\Phi^\lambda_{n-1}  = {\bf R}^{-}_n\vec\Phi^\lambda_n .
\end{eqnarray}
These ladder operators can be determined when we rewrite the tridiagonal
recurrence relation (\ref{triris}) as
\begin{eqnarray}
0 & = &\left[ {{\bf Q}}_{-1,n+1}{\bf R}^{+}_n+{{\bf
Q}}_{0,n}\right]\vec\Phi^\lambda_n+
{{\bf Q}}_{1,n-1}\vec\Phi^\lambda_{n-1}
\end{eqnarray}
and, similarly, as
\begin{eqnarray}
0 & = & {{\bf Q}}_{-1,n+1}\vec\Phi^\lambda_{n+1}+\left[{{\bf
Q}}_{0,n}+{{\bf Q}}_{1,n-1}{\bf R}^{-}_n\right]\vec\Phi^\lambda_{n} .
\end{eqnarray}
Comparing these results with the original definitions (\ref{sbest1})
and shifting the index we find relations for the ladder
operators  ${\bf R}^\pm_n$ themselves:
\begin{eqnarray}
\label{sbest}
{\bf R}^{\mp}_{n}=-\left[ {{\bf Q}}_{\pm 1,n\mp 2}{\bf R}^{\mp}_{n\mp 1}+
{{\bf Q}}_{0,n \mp 1}\right]^{-1}{{\bf Q}}_{\mp 1,n} .
\end{eqnarray}
Again it is sufficient for our purpose
to solve the tridiagonal recurrence relation (\ref{triris}) for the
case $n=0$:
\begin{eqnarray}
0 & = & \left[{{\bf Q}}_{-1,1}{\bf R}^{+}_0+{{\bf Q}}_{0,0}
+{{\bf Q}}_{1,-1}{\bf R}^{-}_0\right]\vec\Phi^\lambda_0 .
\end{eqnarray}
Repeated application of the operator relations (\ref{sbest})
then yields a tridiagonal matrix  valued continued fraction.
A schematic representation of the iteration method
is presented in Fig. 3.

\section{Formulation of the adjoint problem}

In general, the linear infinitesimal generator ${\cal A}_L$ is not self
adjoint
in the extended state space ${\cal C}$. Therefore it becomes necessary
to define another extended state space ${\cal C}^\dagger$ dual to
${\cal C}$ and to investigate the properties of the adjoint infinitesimal
generator ${\cal A}^\dagger_L$. In order to relate the linearized problem
with its adjoint it turns out that the canonical bilinear form for ordinary
differential equations is not appropriate. In the case of delay differential
equations a modified bilinear form has to be introduced.

\subsection{The bilinear form}

The choice of the canonical bilinear form for delay differential equations
is motivated from the Fredholm alternative.
To this end we consider the inhomogeneous version of (\ref{levol})
\begin{eqnarray}
\label{fred}
\left(\left[{\cal A}_L-\frac{d}{d\xi}\right]\vec q_\xi\right)(\theta) =
\vec \chi_\xi(\theta), \quad -\omega\tau\leq\theta\leq 0
\end{eqnarray}
with a $2\pi$ periodic vector valued function $\vec \chi_\xi(\theta)=\vec
\chi_{\xi+2\pi}(\theta)$.
We try to construct a particular solution $\vec q_\xi(\theta)$ of
(\ref{fred}) by the Floquet ansatz
(\ref{ansatz0}) with $\vec \phi^\lambda_\xi(\theta)=\vec
\phi^\lambda_{\xi+2\pi}(\theta)$.
Inserting the Fourier expansion (\ref{ansatz}) for $\vec
\phi^\lambda_\xi(\theta)$ and a corresponding one for the
inhomogeneity $\vec \chi_\xi(\theta)$,
we obtain
\begin{eqnarray}
\label{inhomgen}
\sum_{n=-\infty}^\infty\left(\left[{\cal A}_L-\lambda-in\right]
\vec \phi^\lambda_n\right)(\theta)e^{in\xi} =
\sum_{n=-\infty}^\infty \vec \chi_n(\theta)e^{in\xi}.
\end{eqnarray}
Taking into account the definition (\ref{ansb}) of the infinitesimal
generator ${\cal A}_L$ in the interval $-\omega\tau\leq\theta < 0$,
we conclude from (\ref{inhomgen}) the general
form of the Fourier component $\vec\phi_n^\lambda(\theta)$:
\begin{eqnarray}
\vec\phi_n^\lambda(\theta)= \vec\phi_n^\lambda(0)e^{(\lambda+in)\theta}
+\int_0^\theta ds e^{(\lambda+ in)(\theta-s)}\vec\chi_n(s).
\end{eqnarray}
Correspondingly (\ref{inhomgen}) determines for
the point $\theta=0$ the yet unknown
initial condition $\vec\phi_n^\lambda(0)$. With the definition
(\ref{ldefin}) it fulfills an inhomogeneous vector valued recurrence relation:
\begin{eqnarray}
& & \sum^\infty_{k=-\infty}\left[{{\bf L}}_{k,
n-k}-\delta_{k,0}(\lambda+i
n){\bf I}\right]\vec\phi^\lambda_{n-k}(0) \nonumber\\
\label{rekursinh}
& & = \vec\chi_n(0)-\sum_{k=-\infty}^\infty
\int_{-\omega\tau}^0d\theta\int_0^\theta dse^{(\lambda+i(n-k))(\theta-s)}
{\bf \Omega}_k(\theta)\vec\chi_{n-k}(s) .
\end{eqnarray}
This result suggests how to introduce both the dual space
${\cal C}^\dagger$ and the bilinear form.
We assume that  ${\cal C}^\dagger$ consists of $n$ dimensional vector
valued functions defined on the interval $[0,\omega\tau]$ and that the
bilinear form is given by
\begin{eqnarray}
\label{bilint}
\left(\vec\psi^{\dagger}_\xi(s),\vec\phi_\xi(\theta)\right)_\xi =
\left<\vec\psi^\dagger_\xi(0),\vec\phi_\xi(0)\right>
-\int^0_{-\omega\tau}d\theta\int^\theta_0ds\left<
\vec\psi^\dagger_\xi(s-\theta),{\bf \Omega}_{\xi+s-\theta}(\theta)
\vec\phi_\xi(s)\right>
\end{eqnarray}
for all $\vec\phi_\xi\in{\cal C}$ and $\vec\psi^\dagger_\xi
\in{\cal C}^\dagger$,
where $<,>$ denotes the usual canonical scalar product. Note that each
delay system and each time periodic reference state induces its own
bilinear form due to (\ref{omdef}). Furthermore we observe that the
explicit time dependent bilinear form (\ref{bilint}) for a time periodic
reference state reduces to the corresponding one for a time independent
reference state \cite{wo}.\\

With the bilinear form (\ref{bilint}) the inhomogeneous recurrence
relation (\ref{rekursinh}) can be rewritten according to
\begin{eqnarray}
\label{rekursinh2}
\sum^\infty_{k=-\infty}\left[{{\bf L}}_{k,
n-k}-\delta_{k,0}(\lambda+i
n){\bf I}\right]\vec\phi^\lambda_{n-k}(0) = \frac{1}{2\pi}\int_0^{2\pi}
d\xi\left({\bf A}^\dagger_{\xi,n}(s),\vec\chi_\xi(\theta)\right)_\xi ,
\end{eqnarray}
where the matrix valued functions ${\bf A}^\dagger_{\xi,n}(s)$
are given by
\begin{eqnarray}
\label{alphadef}
{\bf A}^\dagger_{\xi,n}(s)=e^{-in\xi}e^{-(\lambda+in)s}{\bf I},
\quad 0\leq s\leq\omega\tau .
\end{eqnarray}
Thus we obtain the following Fredholm alternative  for solving the
inhomogeneous equation (\ref{fred}).
If the parameter $\lambda$ does not coincide with a Floquet eigenvalue,
we read off from (\ref{rekursinh2}) that there exists a unique solution.
Otherwise we can only expect a solution if the inhomogeneity
$\vec \chi_\xi$ fulfills a solvability condition which involves the
bilinear form (\ref{bilint}).
This solvability condition will be concretized below after having defined
the adjoint operator ${\cal A}^\dagger_L$
and its corresponding Floquet eigensolutions.

\subsection{The adjoint operator}

The bilinear form (\ref{bilint}) can be applied to describe the evolution
of the linearized delay system also in the dual extended state space
${\cal C}^\dagger$. To this end we require that the bilinear form between the
state vector $\vec q_\xi\in{\cal C}$ and its dual
$\vec q^{\,\dagger}_\xi\in{\cal C}^\dagger$ becomes time independent:
\begin{eqnarray}
\label{zeitun}
0  & = &\frac{d}{d\xi}\left(\vec q^{\,\dagger}_\xi(s),
\vec q_\xi(\theta)\right)_\xi .
\end{eqnarray}
As the bilinear form (\ref{bilint}) does explicitly depend on the
time $\xi$ via the matrix valued density ${\bf \Omega}_\xi(\theta)$,
we derive from (\ref{levol}) and (\ref{zeitun}) the evolution equation
in ${\cal C}^\dagger$
\begin{eqnarray}
\label{adjulevol}
\frac{d}{d\xi}\vec q^{\,\dagger}_\xi(s) & = & -\left({\cal A}^\dagger_L
\vec q^{\,\dagger}_\xi\right)(s),\quad 0\leq s\leq\omega\tau ,
\end{eqnarray}
where the adjoint infinitesimal generator ${\cal A}^\dagger_L$ obeys:
\begin{eqnarray}
\label{adju0}
\left({\cal A}^\dagger_L \vec q^{\,\dagger}_\xi,\vec q_\xi\right)_\xi
 =  \left(\vec q^{\,\dagger}_\xi,{\cal A}_L \vec q_\xi\right)_\xi
-\int^0_{-\omega\tau}d\theta\int^\theta_0ds\left<\vec q^{\,\dagger}_\xi
(s-\theta),\frac{\partial}{\partial \xi}{\bf\Omega}_{\xi+s-\theta}(\theta)
\vec q_\xi(s)\right> .
\end{eqnarray}
When we use the definition (\ref{ansb}) of the infinitesimal operator
${\cal A}_L$ we obtain from (\ref{adju0}) after a partial integration the
following expression for the adjoint infinitesimal generator
${\cal A}^\dagger_L$:
\begin{eqnarray}
\label{adju}
\left({\cal A}^\dagger_L \vec q^{\,\dagger}_\xi\right)(s)=
& = & \left\{\begin{array}{cc}\displaystyle -\frac{d}{ds}\vec
q^{\,\dagger}_\xi(s) , & 0<s\leq\omega\tau , \\ \displaystyle
\int^{\omega\tau}_0ds'\vec q^{\,\dagger}_\xi(s') {\bf\Omega}_{\xi+s'}(-s'),
& s=0. \end{array}\right.
\end{eqnarray}

\subsection{The adjoint recurrence relation}

We are now in the position to solve the adjoint problem defined by
(\ref{adjulevol}) and (\ref{adju}).
In close analogy to the procedure in Section III.C we perform the
Floquet ansatz
\begin{eqnarray}
\vec q^{\,\dagger}_\xi(s)=e^{-\lambda \xi}\vec\psi^{\dagger\lambda}_\xi(s)
\end{eqnarray}
with the $2\pi$ periodic adjoint Floquet eigensolution
\begin{eqnarray}
\label{psidef1}
\vec\psi^{\dagger\lambda}_\xi(s)=\sum_j\vec\psi^{\dagger\lambda}_j(s)
e^{-i j\xi} .
\end{eqnarray}
Evaluating (\ref{adjulevol}) and (\ref{adju}) in the interval
$0 < s\leq\omega\tau  $ fixes the form of the Fourier components according to
\begin{eqnarray}
\label{psidef2}
\vec\psi^{\dagger\lambda}_j(s)=\vec\psi^{\dagger\lambda}_je^{-(\lambda+i j)s},
\end{eqnarray}
whereas the case $s=0$ leads to the corresponding homogeneous
vector valued recurrence relation
\begin{eqnarray}
\label{rekursad}
0=\sum^\infty_{k=-\infty}\vec\psi^{\dagger\lambda}_{j+k}
\left[{{\bf L}}_{k,j}-\delta_{k,0}(\lambda+i j){\bf I}\right] .
\end{eqnarray}
In order to solve (\ref{rekursad}) for the Fourier components
$\vec\psi^{\dagger\lambda}_{j}$ of the adjoint Floquet eigensolutions
and the respective Floquet eigenvalues $\lambda$, we proceed along
similar lines as in Section IV.A. First we define adjoint ladder
operators ${\bf Z}^m_j$ with arbitrary $m$ and the identity
${\bf Z}^0_j = {\bf I}$ in analogy to (\ref{s1}):
\begin{eqnarray}
\label{adla}
\vec\psi^{\dagger\lambda}_{j+m} =
\vec\psi^{\dagger\lambda}_{j} {\bf Z}^m_j \, .
\end{eqnarray}
Inserting (\ref{adla}) in the homogeneous vector valued recurrence
relation (\ref{rekursad}) we then obtain a corresponding one for
the adjoint ladder operators:
\begin{eqnarray}
\label{rekursadla}
0=\sum^\infty_{k=-\infty} {\bf Z}^k_j
\left[{{\bf L}}_{k,j}-\delta_{k,0}(\lambda+i j){\bf I}\right] .
\end{eqnarray}
A careful comparison between (\ref{rekurssmn}) and (\ref{rekursadla})
reveals that the recurrence relations for the ladder operators ${\bf S}^m_n$
and their adjoint ${\bf Z}^m_j$ are not independent from each other.
Indeed, they are mapped into each other by the prescription
\begin{equation}
\label{pre}
{\bf L}_{k,n-k} {\bf S}^{-k}_n = {\bf Z}^{-k}_n {\bf L}_{-k,n} \, .
\end{equation}
This means that the adjoint ladder operators ${\bf Z}^m_j$ can be
immediately calculated, once the ladder operators ${\bf S}^m_n$ are
known. However, this does not imply that the solution of the adjoint
problem directly follows from the linear problem. Iteratively inserting
the operator recurrence relation (\ref{rekursadla})
in the vector valued recurrence relation (\ref{rekursad}) for $j = 0$
yields with (\ref{pre})
\begin{equation}
\vec\psi^{\dagger\lambda}_{0} {\bf M} ( \lambda ) = 0 \, .
\end{equation}
Thus the adjoint problem leads to the same condition (\ref{cond}) for the
Floquet eigenvalues $\lambda$, but the Fourier component
$\vec\psi^{\dagger\lambda}_{0}$ has to be determined independently from
the Fourier component $\vec\phi^\lambda_0$ defined by (\ref{homogeneous}).\\

With these definitions we are now able to concretize the Fredholm condition
for solving the inhomogeneous equation (\ref{fred}).
Multiplying (\ref{rekursinh2}) from the left with
$\vec\psi^{\,\dagger\lambda}_n$, performing the summation over all $n$
and taking into account the homogeneous vector valued recurrence relation
(\ref{rekursad}), we yield
\begin{eqnarray}
\frac{1}{2\pi}\int_0^{2\pi}d\xi\left(\sum^\infty_{n=-\infty}
\vec\psi^{\,\dagger\lambda}_n {\bf A}_{\xi,n}(s),\vec\chi_\xi(\theta)
\right)_\xi =0.
\end{eqnarray}
Due to (\ref{alphadef}), (\ref{psidef1}), (\ref{psidef2}) this solvability
condition takes the concise form
\begin{eqnarray}
\frac{1}{2\pi}\int_0^{2\pi}d\xi\left(\vec\psi^{\,\dagger\lambda}_\xi,
\vec\chi_\xi(\theta)\right)_\xi =0.
\end{eqnarray}
Only in the special case that all quantities do not explicitly depend on the
time $\xi$ this reduces to the usual Fredholm condition, i.e. the inhomogeneity
$\vec\chi_\xi$ must be orthogonal to the respective eigensolution of the
adjoint operator ${\cal A}_L^\dagger$ \cite{wo}.

\subsection{Biorthonormality relations}

Finally we show that the Floquet eigensolutions $\vec\phi^\lambda_\xi$
and  $\vec\psi^{\,\dagger\lambda}_\xi$ of the infinitesimal generator
${\cal A}_L$ and its dual  ${\cal A}_L^\dagger$, respectively, can be
chosen to form a biorthonormal set in ${\cal C}$ and ${\cal C}^\dagger$.
First we derive the biorthogonality condition for different Floquet
eigenvalues$\mu\neq\lambda$.
Using the explicit expression for the bilinear form (\ref{bilint})
and applying
our previous results (\ref{fourmeg})--(\ref{ansatz}) and
(\ref{psidef1}), (\ref{psidef2}) we obtain
\begin{eqnarray}
\label{o1}
\left(\vec\psi^{\dagger\lambda}_\xi(s),\vec\phi^\mu_\xi(\theta)\right)_\xi=
\sum_{n,j=-\infty}^\infty\Bigg\{\left<\vec\psi^{\dagger\lambda}_j,
\vec\phi^\mu_n\right>e^{i(n-j)\xi} \hspace*{2cm} \nonumber\\
- \sum_{k=-\infty}^\infty e^{i(n+k-j)\xi}\int_{-\omega\tau}^0d\theta
\int_0^\theta ds \left<\vec\psi^{\dagger\lambda}_j,{\bf \Omega}_k(\theta)
\vec\phi^\mu_n\right>e^{(\lambda+i(j-k))\theta}e^{(\mu-\lambda+i(n+k-j))s}
\Bigg\}.
\end{eqnarray}
An integration with respect to $s$ yields for the second term on the
right hand side
\begin{eqnarray}
-\sum_{n,j,k=-\infty}^\infty e^{i(n+k-j)\xi}\int_{-\omega\tau}^0d\theta
\left<\vec\psi^{\dagger\lambda}_j,{\bf \Omega}_k(\theta)\vec\phi^\mu_n\right>
\frac{e^{(\mu+in)\theta}-e^{(\lambda+i(j-k))\theta}}{\mu-\lambda+i(n+k-j)}.
\end{eqnarray}
With the definition (\ref{ldefin}) of the matrices ${\bf L}_{k,n}$ this
reduces to
\begin{eqnarray}
-\sum_{n,j,k=-\infty}^\infty e^{i(n+k-j)\xi}\frac{\left<\vec\psi^{\dagger
\lambda}_j,\left[{\bf L}_{k,n}-{\bf L}_{k,j-k}\right]
\vec\phi^\mu_n\right>}{\mu-\lambda+i(n+k-j)} ,
\end{eqnarray}
so that we yield from the homogeneous vector valued recurrence relations
(\ref{rekurs}), (\ref{rekursad})
\begin{eqnarray}
\label{o4}
-\sum_{n,j=-\infty}^\infty\left<\vec\psi^{\dagger\lambda}_j,
\vec\phi^\mu_n\right>e^{i(n-j)\xi} .
\end{eqnarray}
>From (\ref{o1})--(\ref{o4}) we conclude the biorthogonality for
$\mu\neq\lambda$:
\begin{eqnarray}
\label{ortho}
\left(\vec\psi^{\dagger\lambda}_\xi(s),\vec\phi^\mu_\xi(\theta)\right)_\xi=0.
\end{eqnarray}
To normalize the biorthogonal set of Floquet eigenfunctions, we introduce
a proper normalization constant in a symmetric way:
\begin{eqnarray}
\vec\phi^\lambda_n=N_\lambda\vec\Phi^\lambda_n,\quad
\vec\psi^{\dagger\lambda}_n=N_\lambda\vec\Psi^{\dagger\lambda}_n .
\end{eqnarray}
>From the requirement
\begin{eqnarray}
\label{requ}
\left(\vec\psi^{\dagger\lambda}_\xi(s),\vec\phi^\lambda_\xi(\theta)
\right)_\xi=1
\end{eqnarray}
we then determine the normalization constant $N_\lambda$
by performing similar calculations as above:
\begin{eqnarray}
N_\lambda=\left[\sum_{n,j=-\infty}^\infty\left<\vec
\Psi^{\dagger\lambda}_j,\left(\delta_{j,n}-\int_{-\omega\tau}^0
d\theta \, \theta e^{(\lambda+i n)\theta}{\bf \Omega}_{j-n}(\theta)\right)
\vec\Phi^\lambda_n\right>\right]^{-1/2}.
\end{eqnarray}
As expected the normalization constant $N_\lambda$ does not explicitly
depend on the time $\xi$.
Summarizing the results (\ref{ortho}) and (\ref{requ}),
the biorthonormality relation reads
\begin{eqnarray}
\label{biortho}
\left(\vec\psi^{\dagger\lambda}_\xi(s),
\vec\phi^\mu_\xi(\theta)\right)_\xi=\delta_{\mu,\lambda} .
\end{eqnarray}

\section{Summary and Conclusions}

The present paper was devoted to systematically develop a Floquet theory
for delay differential equations. At first we approximately determined
a time periodic reference state by extending two standard methods for
ordinary differential equations, namely the Poincar\'e Lindstedt and the
Shohat expansion. Then we tested the stability of this reference state
by constructing Floquet eigensolutions and their corresponding
eigenvalues from matrix valued continued fractions. Finally the Floquet
theory was completed by studying the adjoint problem. The applicability
of our Floquet theory was demonstrated in \cite{chrisi2}.
In particular our analytical treatment provides means of understanding
the mechanism of the continuous control of chaos by self controlling
feedback \cite{pyr1,pyr2}. Previous
investigations have indicated that it becomes crucial to decide whether
an observed stabilized limit cycle corresponds to an unstable cycle of
the system or is produced by the control mechanism
itself \cite{chrisi1,chrisi2}.\\

As the Floquet theory represents a linear stability analysis for a time
periodic reference state there still remains the nonlinear problem to
construct the normal form for an emerging
instability. We expect that this problem
can be tackled in a similar way as in \cite{wo} where synergetic methods
\cite{haken,haken2} are extended to investigate delay differential equations
in the local neighborhood of a time independent reference state. Also
close to the instability of a time periodic reference state the inherent time
scale hierarchy should allow to adiabatically eliminate the fast modes
by using projectors which are induced by the bilinear form
(\ref{bilint}) of the linear stability analysis.
As in \cite{wo} the resulting order parameter equations for the slow modes
should turn out to be of the form of ordinary differential equations.
We stress that the normal form theory is indispensable for classifying
the instabilities of time periodic reference states. Whereas the linear
stability analysis is sufficient to identify the instabilities of
time independent reference states, this is no longer true for time periodic
ones \cite{haken2}.\\

\end{document}